\documentclass[prl,superscriptaddress,amsmath,amssymb,floatfix,twocolumn]{revtex4-1}
\usepackage{graphicx,color}
\usepackage{amsmath,amsfonts,amssymb}
\usepackage{float}
\usepackage[colorlinks,linkcolor=blue,anchorcolor=blue,urlcolor=blue,citecolor=blue]{hyperref}

\begin{document}

\title{Topological sound pumping of zero-dimensional bound states}

\author{Penglin Gao}
\affiliation{Department of Physics, Universidad Carlos III de Madrid, ES-28916 Legan\`es, Madrid, Spain}
\author{Johan Christensen}
\email[Corresponding author.\\]{johan.christensen@uc3m.es}
\affiliation{Department of Physics, Universidad Carlos III de Madrid, ES-28916 Legan\`es, Madrid, Spain}
\date{\today}

\begin{abstract}
{Topological phases have spurred unprecedented abilities for sound, light and matter engineering and recent progress has shown how waves not only confine at the interfaces between topologically distinct insulators, but in the form of zero-dimensional non-propagating states bound to defects or corners. Majorana-like bound states have recently been observed in man-made Kekul\'e textured lattices. We show here how the acoustic version of the associated Jackiw-Rossi vortex embodies a Thouless pumping process, in which the spectral flow of corner states adiabatically merge with the said Majorana-like state. Moreover, we argue how the chirality of the Kekul\'e vortex additionally maps into a 2D quantum-Hall system comprising spatially separated sonic hotspots. We foresee that our findings should provide novel exotic tools to enable contemporary control over  sound.}
\end{abstract}

\maketitle

Bloch bands in man-made crystals not only display forbidden and allowed regions of wave propagation, but also embody the geometrical topology of eigenstates \cite{RevModPhys.83.1057, zhangxj2018topological}. Moreover, topological bulk bands manifest in surface and interface related properties through the notion of the bulk-edge correspondence predicting boundary states via topological invariants. These prominent attributes are responsible for a wealth of thriving frontier investigations unveiling exotic defect-immune guiding of sound and mechanical vibrations \cite{PhysRevLett.100.013904, yangzj2015topological, he2016acoustic, susstrunk2015observation, PhysRevApplied.9.034032,miniaci2018experimental, zhangzw2018directional, PhysRevApplied.12.034014, PhysRevB.101.020301}. Most recently, phononic and photonic higher-order topological insulators (HOTIs) that sustain zero-dimensional (0D) corner states in 2D and 3D systems, have been proposed as a counterpart to topologically protected systems abiding by the bulk-edge correspondence. Along this direction, many fascinating HOTI experiments have been constructed \cite{benalcazar2017quantized,serra2018observation, xue2019acoustic, PhysRevLett.122.204301, el2019corner, PhysRevLett.122.233903, mittal2019photonic}, among which a deep-suwavelength topological lens was proposed capable of breaking the sonic diffraction limit  \cite{zhang2019deep}.\\
In contrast to HOTIs that support lower-dimensional states at their external boundaries, the Jackiw-Rossi vortex has shown capable to bind an exact zero mode within the lattice-bulk \cite{jackiw1981zero, PhysRevLett.98.186809, PhysRevB.82.115120}. Quantum mechanical systems in particular hold great promise for these Majorana-zero modes as they appear highly promising for braiding-based topological quantum computation \cite{PhysRevLett.100.096407, PhysRevB.82.144513, RevModPhys.80.1083}. Recently, a wealth of experimental efforts has sparked curiosity among phononic and photonic researchers in the pursuit of a classical analogy \cite{PhysRevLett.123.196601, chen2019mechanical, PhysRevLett.117.073901, gao2019dirac}. Based on a so-called Kekul\'e binding mechanism in artificially man-made macroscopic lattices, equivalent zero modes of vibrations, sound or light have shown intriguing properties for topological robust single-mode wave control and confinement. Interestingly, it has been shown that such classical implementations of the Jackiw-Rossi binding mechanism facilitate particle-hole symmetry counterparts, however, in the absence of the peculiar Majorana self-conjugation relation.\\ 
In this Letter, we present the acoustic Jackiw-Rossi vortex in the framework of a topological Thouless pumping process. A topological pump constitutes a quantized charge transport in a periodically modulated potential, which is realized when the system parameters are adiabatically and cyclically varied to facilitate topologically robust energy transfer, as has been shown in recent exciting wave-based experiments \cite{PhysRevB.27.6083,PhysRevLett.109.106402,PhysRevLett.120.120501,PhysRevLett.123.034301}. {\color{black} These implemented topologial pumps typically incorporate adiabatic modulation of the geometrical parameters, which leads to a smooth crossing of propagating edge states from one boundary to the opposite. Moreover,  since corner states can be understood through the vectorial Zak phase, it is not surprising that a 2D pumping process can push a corner state excitation into the opposite corner through bulk-mode hybridization \cite{4Dpumping}. Latest experimental implementations of the Jackiw-Rossi vortex employ a Kekul\'e texture to artificially decorated lattices that embody a winding process \cite{PhysRevLett.123.196601, chen2019mechanical}. Here, we show that the binding mechanism indeed can be presented in terms of a pumping process encompassing the polar paramaters of the Kekul\'e modulation, which leads to a sonic spectral flow during the adiabatic structural tuning. Surprisingly, we find that the 0D states in finite crystals evolving from the adiabatic pumping process incorporate corner states that spectrally flow in the form of the coveted Majorana-like zero modes. Simply put, as opposed to the aforementioned findings reporting on the spatial crossing of either edge or corner states, our reported evolution deals with the merging of corner states into a non-propagating state that is bound to a topological defect.} Incorporating fractional and segmentations to the angular phase texture, further corroborates the pumping origin of our topologically robust sonic state. Finally, we show the ability to engineer a distinct chirality as one desires, in which the Kekul\'e angular phase represents a synthetic dimension of one-way flowing sound. In other words, the topological pumping process of the Jackiw-Rossi vortex maps into a 2D quantum-Hall system comprising the well known spatial separation of charges \cite{PhysRevLett.102.187001}, in our case, according to the vortex handedness, we obtain opposing heart-shaped sonic spots. \\
\begin{figure*}
	\centering
	\includegraphics[scale=1]{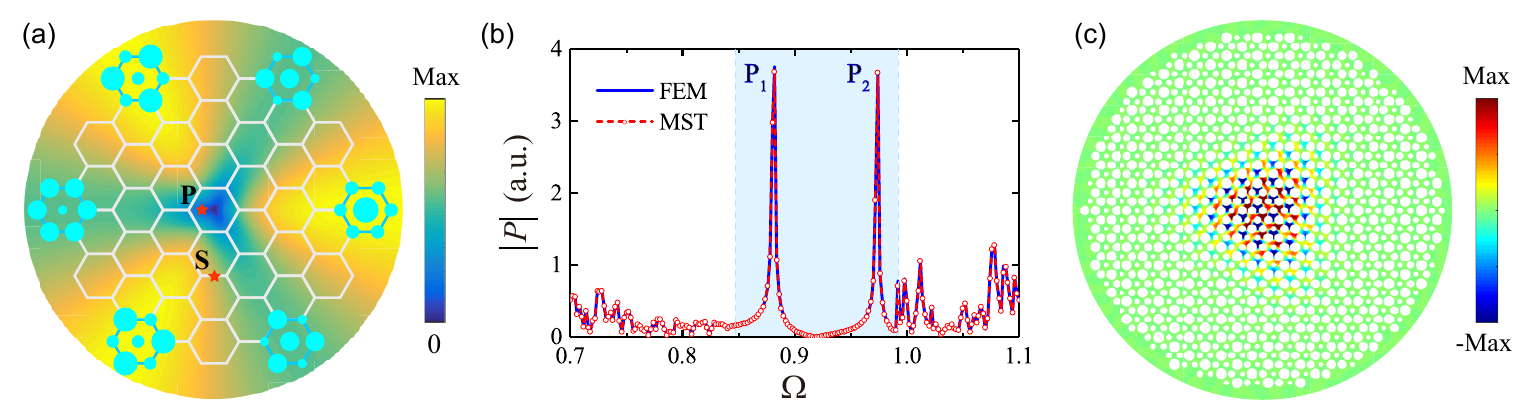}
	\caption{A sonic Jackiw-Rossi vortex hosting a Majorana-like zero mode at its core. (a) Schematic of the Kekel\'e distorted lattice ($n=1$), which illustrates the position dependent crystal variation. The background color indicates the width of the spatial distribution of the Kekel\'e induced band gap, which is exactly zero at the vortex core. The labels $S$ and $P$ indicate the excitation and probe points, respectively. (b) Comparison between the calculated pressure spectra $|P(\Omega)|$ using both FEM and MST predictions. Frequency $\omega$ is normalized to $\Omega=\omega a/2\pi c$, with $c$ being the speed of sound and $a$ the lattice period. (c) The pressure field of the Majorana-like bound state [see $P_1$ in panel (b)] as obtained by multiple scattering simulations.}
	\label{fig1}
\end{figure*}
We begin the study by briefly revisiting the engineering of a Majorana-like zero mode in a distorted sonic lattice by means of a multiple scattering theory (MST). In Fig. \ref{fig1}(a) we depict the acoustic analogue of a Jackiw-Rossi vortex containing a man-made Kekul\'e texture. As recently experimentally verified in Ref. \onlinecite{PhysRevLett.123.196601}, such vortex is readily designed by tuning the acoustically rigid cylinder-radii in a position-dependent fashion, i.e.,
\begin{equation} 
R(\textbf{r})=R_0+\delta R(r) \cos\left[\textbf{K}\cdot\textbf{r} + \phi(\textbf{r})\right],
\end{equation}
where $R_0$ is the undisturbed radius, $\delta R(r)$ denotes the radius variation, $\textbf{K}$ and $\phi(\textbf{r})$ stand for the Kekul\'e wave vector and phase, respectively. The two position-dependent terms read: $\delta R(r)=\Delta \tanh (r/\xi)$ and $\phi(\textbf{r})=n \theta$, with $\xi$ and $n$ being the vortex radius and winding number, respectively. This Kekul\'e texture is indeed a topological winding process that encompasses the gapless core of the Jackiw-Rossi vortex, at which the triangular lattice remains undistorted, i.e., $\delta R(0)=0$. However, the targeted coiling $\phi(\textbf{r})$ determines the angular variation of the width that separates the acoustic valleys at the $\Gamma$ point of the band diagram.  The coloured angular fingerprints of the bandwidth of this band gap, is clearly seen in Fig. \ref{fig1}(a). Beyond the man-made winding, the Kekul\'e texture also encompasses a radial component $\delta R(r)$ [Eq. (1)] that is the responsible actor controlling the tightness of the confined Majorana-like state. We employ a MST in order to predict the complex acoustic interplay among the rigid cylinders of the Jackiw-Rossi vortex in response to a point source \cite{suppl}. In the simulations we selected the following geometrical parameters: $R_0=0.35d$, $\Delta=0.15d$, $\xi=2a$, and $n=1$, with the nearest neighbour spacing $d=a/\sqrt{3}$. Fig. \ref{fig1}(b) depicts the calculated pressure spectra evaluated at point $P$ of the schematic, illustrating a remarkable agreement with FEM predictions. The winding-induced anisotropy of the bandgap that we discussed in Fig. \ref{fig1}(a) is further corroborated through the angled pressure state profile seen in Fig. \ref{fig1}(c), which was also calculated with the MST. Additional robustness studies and experimental verifications of this topological bound state were conducted in ref. \onlinecite{PhysRevLett.123.196601}.\\
\begin{figure*}
	\centering
	\includegraphics[scale=1]{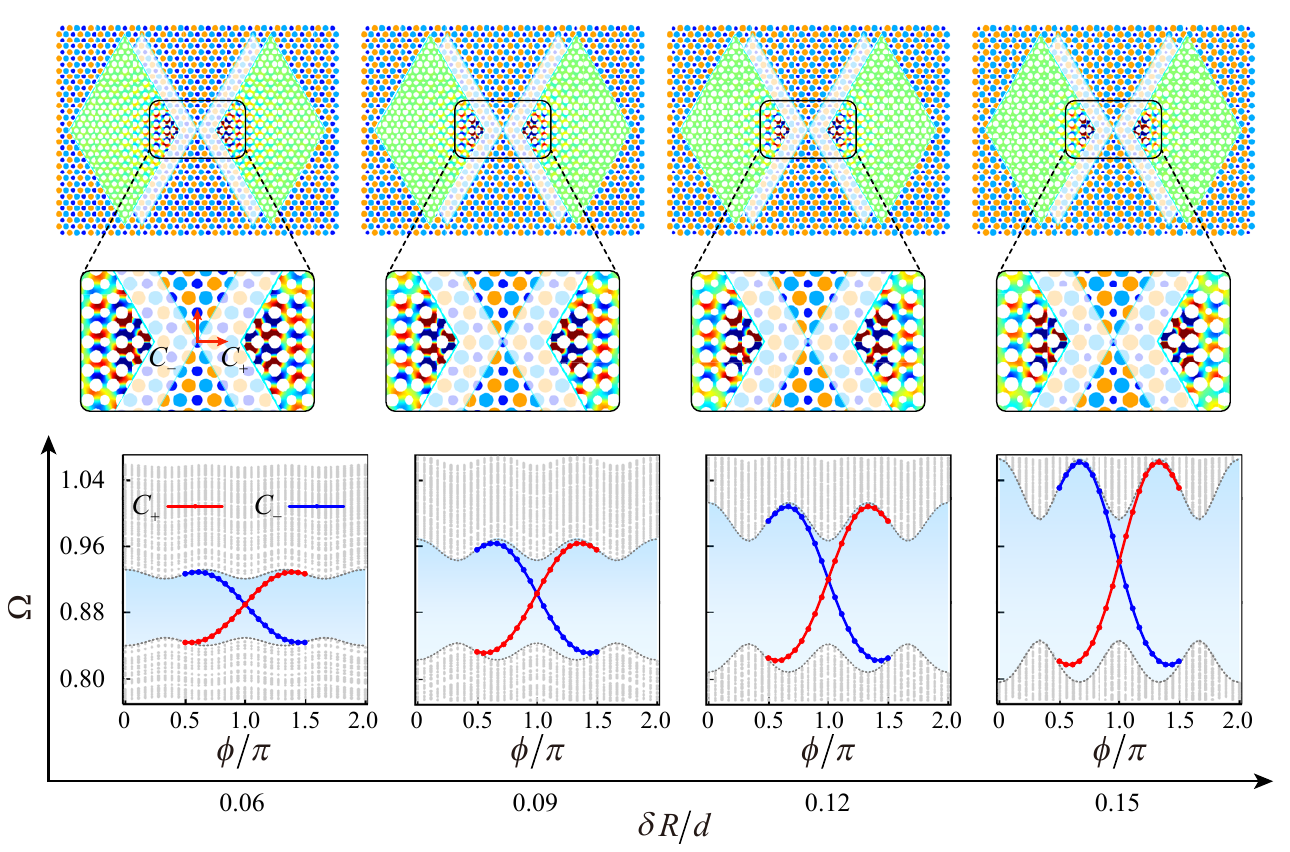}
	\caption{Topological sound pumping of 0D bound states. Upper panels: The background periodic lattice depicts a uniformly Kekul\'e modulated structure with $\phi=\pi$ at various values of $\delta R$ according to the lower panel. Centered at $ \textbf{r} =0$ of the red coordinate system, we cut two rhombi that are capable to sustain topological corner states denoted by $C_{\pm}$. The magnifications of the rhombi terminations illustrate the same topology surrounding the center-cylinder. The cyan border represents a hard-walled interface. Lower panel: Spectral flow of the in-gap states when the  Kekul\'e phase $\phi$ winds a full $2\pi$ period for four selected values of $\delta R$, depicting the various manifestations of the bulk-corner correspondence. When $\delta R \rightarrow 0$ the bound states within the shrinking bandgap approach the Dirac frequency $\Omega_{D}=0.88$.}
	\label{fig2}
\end{figure*}
In what follows, we argue that the sonic Kekul\'e winding indeed embodies an adiabatic pumping process in dependence of the polar Kekul\'e phase. As we discussed earlier, similar to the quantum Thouless pumping describing an adiabatic variation of charges that are quantized as a topological invariant, we now demonstrate how the Majorana-like zero mode adiabatically evolves from the spectral pumping of corner states. The polar Kekul\'e parameters include both $\phi(\textbf{r})$ and  $\delta R(r)$ that will serve as the corner stone behind the pumping process. To cast this process within the framework of topological corner states and the afore discussed Majorana-like states, we define the spatial origin of such 0D modes at the center-cylinder of their triangular lattices. In doing this, we design the same geometrical topology for the system depicted in Fig. \ref{fig1}(a) and the finite rhombi in Fig. \ref{fig2}. To elaborate on the significance of this approach, we numerically compute the spectral flow of the 0D in-gap states of finite rhombi made of the previously discussed triangular lattice of rigid cylinders. The spectral flow maps the eigenvalues of a sonic rhombus as a function of the Kekul\'e phase $\phi$. These computations depict how the 0D modes emerge from the bulk-bands in dependence of the angular Kekul\'e winding in the form of two in-gap residing corner states, denoted by $C_{\pm}$ as seen in Fig. \ref{fig2}. The $\pm$ indicates that that two corner states contain opposing chirality during the pumping process, which we will elaborate later. The adiabatic pumping process gathers both Majorana-like and corner states under the same umbrella in that we further must consider the radial component $\delta R(r)$ of the cylinder radius modulation. Thanks to its hyperbolic tangential component, as explained in the former, the radial variation approaches zero at the cluster core in Fig. \ref{fig1}(a), at which the zero mode is pinned spatially. The spectral flows of the corner states that are presented in Fig. \ref{fig2} are computed against $\delta R$ to underline the second dimension of the adiabatic geometrical variation. Moving from right to left, i.e., when decreasing $\delta R$ we immediately predict a narrowing of the topological bandgap, which results in a smooth shift comprising the spectral winding of corner states $C_{\pm}$ towards the Dirac frequency $\Omega_{D}=0.88$, at which the Majorana-like bound state is fixed. In the upper panel of Fig. \ref{fig2}, for the degenerate spectral flow of corner states ($\phi=\pi$), we visualize these 0D states that live at opposite corners, here shown for different values of $\delta R$. For these particular geometrical parameters we depict the uniform triangular lattices in the background from which the rhombi are cropped out. The vortex core is thus understood as an collective formation of corner states $C_{-}$ ($C_{+}$) that wind in a clockwise (counter-clockwise) manner with $n=-1$ ($n=1$) through the adiabatic parameter $\phi(\textbf{r})$. Thus the corner states, as seen in the spectral flows, possess a handedness where the vortex and antivortex come in pairs. Interestingly, one could consider the Kekul\'e winding parameter $\phi$ as the third coordinate via a dimensional extension. To this end, the spectral flow maps exactly into the gapless dispersion relation of chiral hinge states in $C_{2z}T$ invariant 3D topological crystalline insulators \cite{lee2019fractional, PhysRevB.96.245115}. In other words, one-way polarized confined states along the hinges of HOTIs find their counterpart in our Kekul\'e texture that adiabatically pumps the $C_{\pm}$ states across the bulk and the bandgap with respect to their chiralities.\\
\begin{figure}
	\centering
	\includegraphics[width=1.0\columnwidth]{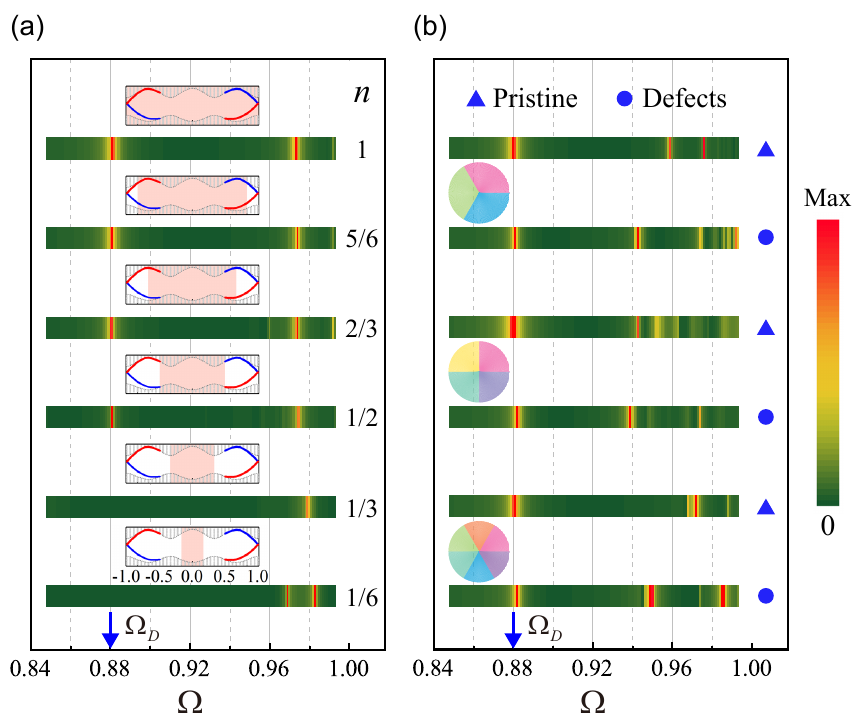}
	\caption{Modulation of the angular Kekul\'e pumping parameter $\phi(\textbf{r})$. (a) Computed pressure spectra $|P(\Omega)|$ within the topological bandgap for vortices of fractional winding $n$. The division of the $2\pi$ Kekul\'e cycle of the spectral flows is indicated by the shaded pink background. (b) Same as before, now the vortex is segmented by $N$ uniform partitions of equal Kekul\'e phase increments for $n=1$, as illustrated by the pie charts. An additional spectrum is computed comprising cylinder-radii perturbations with $\delta R=0$ within a defect zone of radius $R_{D}=3d$.}
	\label{fig3}
\end{figure}
To additionally provide evidence that the Jackiw-Rossi vortex constitutes a topological pumping process, we discuss how an incomplete $2\pi$ Kekul\'e phase $\phi(\textbf{r})$ can suffice while it still maintains the plenary adiabatic modulation. To address this effect, two groups of simulations are performed for vortices of fractional winding numbers [see Fig. \ref{fig3}(a)] and containing $N$ uniform partitions [see Fig. \ref{fig3}(b)]. Except for $\phi(\textbf{r})$, all parameters remain as in Fig. \ref{fig1}. The fractional vortices in Fig. \ref{fig3}(a) contain a continuous but incomplete Kekul\'e phase modulation, which does not span over the entire $2\pi$ range. For the fractional vortices, the response spectra depicted in Fig. \ref{fig3}(a) clearly reveal their connection to the winding number $n$. By gradually breaking the integer winding number into fractions, we see that the bound state at $\Omega_{D}$ gradually shrinks down to the critical point at $n=1/2$, beyond which the state ceases to exist. To unravel the significance of this threshold, we must return to the former study that exemplified how topological pumping depicts an evolution process of corner states ruled by the spectral flow. These states that confine within the bounds of the Kekul\'e phase $\pi/2 \leq|\phi(\textbf{r})|\leq \pi$, which are depicted in Fig. \ref{fig3}(a), must self-similarly reside within a nontrivial phase-zone (pink backgrund) as controlled by the structuring of $n$. Hence, for $n <1/2$ the adiabatic winding does not acquire a sufficiently large angular phase accumulation to enter the spectral flow of corner states, which strongly corroborates that the Majorana-like state originates from a topological pumping process. Next, as shown in Fig. \ref{fig3}(b), we conduct an additional study where the angular variation of $\phi(\textbf{r})$ is segmented into $N$ ($=3,4,6$) uniform partitions of equal phase intervals $2\pi/N$. Instead of fractioning the winding as we did in the former case, the present analysis constitutes a nonsmooth winding ($n=1$) with respect to the said rough segmentations. Intuitively, one would expect poor binding of topological vortices of few segments, but not only do we predict the usual bound state at $\Omega_{D}$ down to $N=3$, it is further seen in Fig. \ref{fig3}(b) that symmetry-preserving defects have close to no influence on the topologically protected pressure peak $|P(\Omega)|$. Conclusively, apart from multiple additional features appearing in the spectrum when defects are added [Fig. \ref{fig3}(b)], once more it is confirmed that the Majorana-like states originate from a rather versatile winding implementation.\\
\begin{figure}
	\centering
	\includegraphics[scale=1]{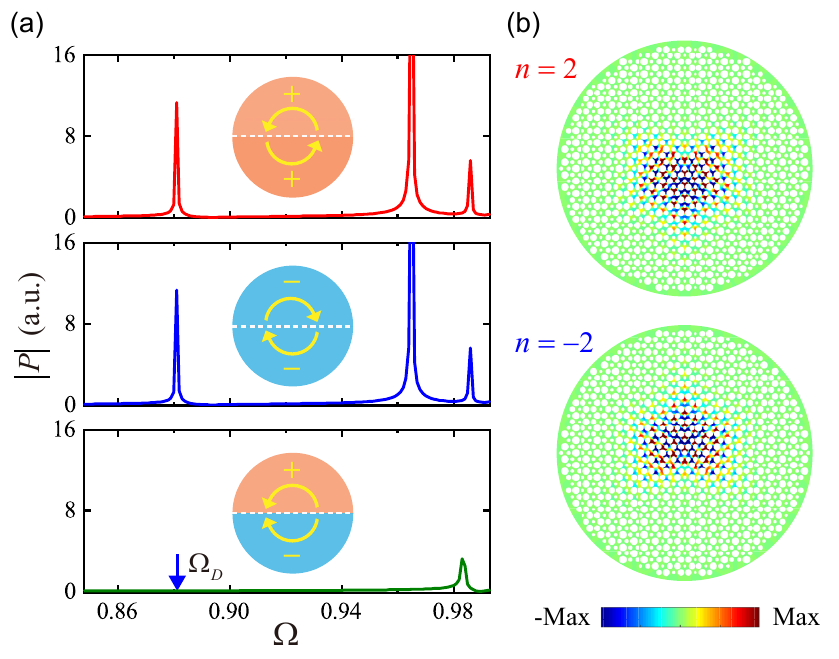}
	\caption{Chiral engineering: we coalesce two semi-clusters each hosting a vortex of particular handedness. Specifically, we combine two vortices of right-handed ($n = 2 = 1+1$) and left-handed ($n = -2 = -1-1$) chirality. In addition, we construct a cluster constituting a combined vortex and antivortex. (a) Plots depict the corresponding pressure spectra $|P(\Omega)|$. As indicated by the arrows in the insets, the symbols ``$-$" and ``$+$" label, respectively, a clockwise and counter-clockwise $2\pi$ phase winding in the corresponding semi-clusters.  (b) The spatial pressure maps of the two clusters of opposite vortex handedness display a spatial separation of the bound states in the form of opposing heart-shaped hotspots, when excited at $\Omega_{D}$.}
	\label{fig4}
\end{figure}
The adiabatic evolution of the spectrally flowing corner states as we discussed earlier, projects the otherwise time-reversal symmetrical angular Kekul\'e texture $\phi(\textbf{r})$ along the third coordinate of HOTIs. Put differently, the handedness of our man-made Jackiw-Rossi vortices finds it imminent counterpart in topologically protected one-way hinge states flowing along the $k_z$ axis in momentum space. To shed more light on this similarity, we distinctively engineer some vortices to enable the synthetic dimensional extension in question. As rendered in Fig. \ref{fig4}(a) we coalesce two semi-clusters with each having a full $2\pi$ winding. If a vortex and an antivortex are combined, they cancel each other out as seen in the lower panel of this figure and a bound state cannot be formed. From the chiral pumping process discussed in Fig. \ref{fig2}, the spectral flows of two non-propagating corner states were characterized by their respective vortex handedness. In order to probe them individually via the Majorana-like state, we now consider a collective Kekul\'e winding among the two semi-clusters, i.e., $n=2$ or $n=-2$, which in turn display identical pressure spectra as computed in Fig. \ref{fig4}(a). Interestingly, we find that our man-made vortices bear a striking resemblance to the edge states of a 2D Chern insulator where counterflowing chiral edge states are separated into opposing "lanes", in that heart-shaped sonic hotspots are formed that orient with respect to the chirality of the Kekul\'e winding as seen in Fig. \ref{fig4}(b). \\
In this Letter, we have gathered sonic zero-dimensional corner states and Majorana-like bound states within the same framework of a topological Thouless pump. By imposing a so-called Kekul\'e texture to artificial sonic lattices to craft a topological Jackiw-Rossi vortex, we unravel its binding mechanism of a bound state through spectrally flowing   corner states. The chiral nature of the topological pumping process has been further enlightened by incorporating fractional and segmented winding and by arguing how spatially distinct acoustic hotspots are formed, akin to the unidirectional characteristics of a 2D quantum-Hall insulator.
\begin{acknowledgments}
J. C. acknowledges the support from the European Research Council (ERC) through the Starting Grant No. 714577 PHONOMETA and from the MINECO through a Ram\'on y Cajal grant (Grant No. RYC-2015-17156).
\end{acknowledgments}

\bibliography{references}


\end{document}